\documentclass[magazine]{IEEEtran}

\usepackage[letterpaper, left=0.625in, right=0.625in, bottom=1.04in, top=0.73in, columnsep=0.26in]{geometry}

\usepackage[T1]{fontenc}
\usepackage[utf8]{inputenc}
\usepackage{cite}
\usepackage{amsmath, mathtools}
\usepackage{amsfonts,amsthm,bm}
\usepackage{amssymb}
\usepackage{comment}
\usepackage{graphicx}
\usepackage{standalone}
\usepackage{dsfont}
\usepackage{mathtools}
\usepackage{color, colortbl}
\usepackage{soul}
\usepackage[normalem]{ulem}
\usepackage[acronym,shortcuts]{glossaries}
\usepackage{arydshln}
\usepackage{bbm}
\usepackage{svg} 
\usepackage{algorithm,algorithmic}
\usepackage{adjustbox}
\usepackage[inline]{enumitem}
\usepackage{multirow}
\usepackage{tabularx}
\usepackage{booktabs}

\usepackage{comment}
\usepackage{tikz}
\usepackage{pgfplots}
\pgfplotsset{compat=newest}
\pgfplotsset{plot coordinates/math parser=false}
\newlength\fheight
\newlength\fwidth
\usetikzlibrary{plotmarks,patterns,decorations.pathreplacing,backgrounds,calc,arrows,arrows.meta,spy,matrix}
\usepgfplotslibrary{patchplots,groupplots}
\usepackage{tikzscale}
\usepackage{balance}
\usepackage{stfloats} 

\usepackage{hyperref}
\usepackage[capitalize]{cleveref}
\crefname{section}{Sec.}{Secs.}

\usepackage{subcaption}
\usepackage{graphicx}
\usepackage{caption}
\usepackage{xcolor}
\definecolor{ieeeblue}{HTML}{00629B}
\definecolor{ieeeorange}{HTML}{FFA300}
\definecolor{ieeegreen}{HTML}{78BE20}
\definecolor{ieeered}{HTML}{BA0C2F}

\usetikzlibrary {patterns.meta}

\usetikzlibrary{patterns}

\usepackage{ulem}

\providecolor{added}{rgb}{0,0,1}
\providecolor{deleted}{rgb}{1,0,0}

\makeatletter
\newcommand{\new}[1]{{\textcolor{blue}{#1}}}

\makeatletter
\newcommand\remembertext[2]{
  \immediate\write\@auxout{\unexpanded{\global\long\@namedef{mytext@#1}{#2}}}%
  {\color{blue} #2}%
}
\newcommand\recalltext[1]{%
  \new{\ifcsname mytext@#1\endcsname
    \fontsize{10.5}{12.5}\selectfont\@nameuse{mytext@#1}%
  \else
    ``??''
  \fi
}}
\makeatother

\newacronym{3gpp}{3GPP}{3rd Generation Partnership Project}
\newacronym{adc}{ADC}{Analog to Digital Converter}
\newacronym{pdr}{PDR}{Packet Delivery Ratio}
\newacronym{5g}{5G}{5th generation}
\newacronym{6g}{6G}{6th generation}
\newacronym{ai}{AI}{Artificial Intelligence}
\newacronym{aimd}{AIMD}{Additive Increase Multiplicative Decrease}
\newacronym{am}{AM}{Acknowledged Mode}
\newacronym{tn}{TN}{Terrestrial Network}
\newacronym{amc}{AMC}{Adaptive Modulation and Coding}
\newacronym{aqm}{AQM}{Active Queue Management}
\newacronym{awgn}{AGWN}{Additive White Gaussian Noise}
\newacronym{balia}{BALIA}{Balanced Link Adaptation}
\newacronym{bdp}{BDP}{Bandwidth-Delay Product}
\newacronym{bf}{BF}{beamforming}
\newacronym{uu}{Uu}{Universal um}
\newacronym{refp}{RP}{Reference Point}
\newacronym{cc}{CC}{Congestion Control}
\newacronym{cdf}{CDF}{Cumulative Distribution Function}
\newacronym{cn}{CN}{Core Network}
\newacronym{cqi}{CQI}{Channel Quality Information}
\newacronym{cp}{CP}{Control Plane}
\newacronym{up}{UP}{User Plane}
\newacronym{upf}{UPF}{User Plane Function}
\newacronym{csirs}{CSI-RS}{Channel State Information - Reference Signal}
\newacronym{sib}{SIB}{System Information Block}
\newacronym{dc}{DC}{Dual Connectivity}
\newacronym{rb}{RB}{Resource Block}
\newacronym{dce}{DCE}{Direct Code Execution}
\newacronym{dci}{DCI}{downlink control onformation}
\newacronym{udp}{UDP}{User Datagram Protocol}
\newacronym{dl}{DL}{downlink}
\newacronym{drl}{DRL}{deep reinforcement learning}
\newacronym{fcfs}{FCFS}{first-come-first-served}
\newacronym{dmr}{DMR}{Deadline Miss Ratio}
\newacronym{fspl}{FSPL}{free-space path loss}
\newacronym{dmrs}{DMRS}{DeModulation Reference Signal}
\newacronym{e2e}{E2E}{End-to-End}
\newacronym{ppp}{PPP}{Poission Point Process}
\newacronym{aoi}{AoI}{Area of Interest}
\newacronym{cpu}{CPU}{Central Processing Unit}
\newacronym{gpu}{GPU}{Graphics Processing Unit}
\newacronym{tpu}{TPU}{Tensor Processing Unit}
\newacronym{ta}{TA}{Timing Advance}
\newacronym{si}{SI}{Study Item}
\newacronym{ecn}{ECN}{Explicit Congestion Notification}
\newacronym{edf}{EDF}{Earliest Deadline First}
\newacronym{enb}{eNB}{eNodeB}
\newacronym{epc}{EPC}{Evolved Packet Core}
\newacronym{es}{ES}{Edge Server}
\newacronym{cav}{CAV}{Connected and Autonomous Vehicle}
\newacronym{fdma}{FDMA}{Frequency Division Multiple Access}
\newacronym{fdd}{FDD}{Frequency Division Duplexing}
\newacronym{tdm}{TDM}{Time Division Multiplexing}
\newacronym{upa}{UPA}{Uniform Planar Array}
\newacronym{car}{CAR}{Circular Aperture Reflector }
\newacronym[firstplural=Radio Access Technologies (RATs)]{rat}{RAT}{Radio Access Technology}
\newacronym[firstplural=Radio Access Technology (RTs)]{rt}{RT}{Radio Technology}
\newacronym{fs}{FS}{Fast Switching}
\newacronym{isd}{ISD}{inter-site distance}
\newacronym{ftp}{FTP}{File Transfer Protocol}
\newacronym{gnb}{gNB}{Next Generation NodeB}
\newacronym{harq}{HARQ}{Hybrid Automatic Repeat reQuest}
\newacronym{hetnet}{HetNet}{Heterogeneous Network}
\newacronym{hh}{HH}{Hard Handover}
\newacronym{hol}{HOL}{Head-of-Line}
\newacronym{ia}{IA}{Initial Access}
\newacronym{imt}{IMT}{International Mobile Telecommunication}
\newacronym{iot}{IoT}{Internet of Things}
\newacronym{los}{LOS}{Line of Sight}
\newacronym{lte}{LTE}{Long Term Evolution}
\newacronym{m2m}{M2M}{Machine to Machine}
\newacronym{mac}{MAC}{Medium Access Control}
\newacronym{mc}{MC}{Multi-Connectivity}
\newacronym{mcs}{MCS}{Modulation and Coding Scheme}
\newacronym{mec}{MEC}{Mobile Edge Cloud}
\newacronym{mi}{MI}{Mutual Information}
\newacronym{mimo}{MIMO}{Multiple Input Multiple Output}
\newacronym{mmwave}{mmWave}{millimeter wave}
\newacronym{mptcp}{MP-TCP}{Multipath TCP}
\newacronym{mr}{MR}{Maximum Rate}
\newacronym{mss}{MSS}{Maximum Segment Size}
\newacronym{mtd}{MTD}{Machine-Type Device}
\newacronym{mtu}{MTU}{Maximum Transmission Unit}
\newacronym{nfv}{NFV}{Network Function Virtualization}
\newacronym{vnf}{VNF}{Virtualization Network Function}
\newacronym{gv}{GV}{ground vehicle}
\newacronym{vec}{VEC}{Vehicular Edge Computing}
\newacronym{dn}{DN}{Data Network}
\newacronym{sdn}{SDN}{Software Defined Networking}
\newacronym{nlos}{NLOS}{Non Line of Sight}
\newacronym{nlosb}{NLOSb}{Building Non Line of Sight}
\newacronym{nlosv}{NLOSv}{Vehicle Non Line of Sight}
\newacronym{nr}{NR}{New Radio}
\newacronym{ofdm}{OFDM}{Orthogonal Frequency Division Multiplexing}
\newacronym{pdcch}{PDCCH}{Physical Downlonk Control Channel}
\newacronym{sctp}{SCTP}{Stream Control Transport Protocol}
\newacronym{sdap}{SDAP}{Service Data Adaptation Protocol}
\newacronym{pdcp}{PDCP}{Packet Data Convergence Protocol}
\newacronym{pdsch}{PDSCH}{Physical Downlink Shared Channel}
\newacronym{pdu}{PDU}{Packet Data Unit}
\newacronym{pf}{PF}{Proportional Fair}
\newacronym{pgw}{PGW}{Packet Gateway}
\newacronym{sgw}{SGW}{Serving Gateway}
\newacronym{phy}{PHY}{Physical}
\newacronym{pbch}{PBCH}{Physical Broadcast Channel}
\newacronym[plural=\gls{mme}s,firstplural=Mobility Management Entities (MMEs)]{mme}{MME}{Mobility Management Entity}
\newacronym{prb}{PRB}{Physical Resource Block}
\newacronym{pss}{PSS}{Primary Synchronization Signal}
\newacronym{pucch}{PUCCH}{Physical Uplink Control Channel}
\newacronym{pusch}{PUSCH}{Physical Uplink Shared Channel}
\newacronym{rach}{RACH}{Random Access Channel}
\newacronym{ran}{RAN}{Radio Access Network}
\newacronym{red}{RED}{Random Early Detection}
\newacronym{rf}{RF}{Radio Frequency}
\newacronym{rlc}{RLC}{Radio Link Control}
\newacronym{rlf}{RLF}{Radio Link Failure}
\newacronym{rrc}{RRC}{Radio Resource Control}
\newacronym{rrm}{RRM}{Radio Resource Management}
\newacronym{rr}{RR}{Round Robin}
\newacronym{rs}{RS}{Remote Server}
\newacronym{rsrp}{RSRP}{Reference Signal Received Power}
\newacronym{rss}{RSS}{Received Signal Strength}
\newacronym{rtt}{RTT}{Round Trip Time}
\newacronym{rw}{RW}{Receive Window}
\newacronym{rx}{RX}{Receiver}
\newacronym{sa}{SA}{standalone}
\newacronym{sack}{SACK}{Selective Acknowledgment}
\newacronym{sap}{SAP}{Service Access Point}
\newacronym{sch}{SCH}{Secondary Cell Handover}
\newacronym{scoot}{SCOOT}{Split Cycle Offset Optimization Technique}
\newacronym{sdma}{SDMA}{Spatial Division Multiple Access}
\newacronym{sinr}{SINR}{Signal to Interference plus Noise Ratio}
\newacronym{sm}{SM}{Saturation Mode}
\newacronym{snr}{SNR}{Signal-to-Noise Ratio}
\newacronym{son}{SON}{Self-Organizing Network}
\newacronym{ss}{SS}{Synchronization Signal}
\newacronym{srs}{SRS}{Sounding Reference Signal}
\newacronym{sss}{SSS}{Secondary Synchronization Signal}
\newacronym{tb}{TB}{Transport Block}
\newacronym{tcp}{TCP}{Transmission Control Protocol}
\newacronym{tdd}{TDD}{Time Division Duplexing}
\newacronym{tdma}{TDMA}{Time Division Multiple Access}
\newacronym{tfl}{TfL}{Transport for London}
\newacronym{tm}{TM}{Transparent Mode}
\newacronym{prr}{PRR}{Packet Reception Ratio}
\newacronym{trp}{TRP}{Transmitter Receiver Pair}
\newacronym{tti}{TTI}{Transmission Time Interval}
\newacronym{ttt}{TTT}{Time-to-Trigger}
\newacronym{tx}{TX}{Transmitter}
\newacronym{ue}{UE}{User Equipment}
\newacronym{ul}{UL}{uplink}
\newacronym{uml}{UML}{Unified Modeling Language}
\newacronym{um}{UM}{Unacknowledged Mode}
\newacronym{utc}{UTC}{Urban Traffic Control}
\newacronym{vm}{VM}{Virtual Machine}
\newacronym{rsrq}{RSRQ}{Reference Signal Received Quality}
\newacronym{rssi}{RSSI}{Received Signal Strength Indicator}
\newacronym{crs}{CRS}{Cell Reference Signal}
\newacronym{v2v}{V2V}{Vehicle-to-Vehicle}
\newacronym{v2i}{V2I}{Vehicle-to-Infrastructure}
\newacronym{v2n}{V2N}{Vehicle-to-Network}
\newacronym{v2x}{V2X}{Vehicle-to-Everything}
\newacronym{vn}{VN}{Vehicular Node}
\newacronym{dsrc}{DSRC}{Dedicated Short Range Communication}
\newacronym{ci}{CI}{context information}
\newacronym{voi}{VoI}{value of information}
\newacronym{gps}{GPS}{Global Positioning System}
\newacronym{qos}{QoS}{Quality of Service}
\newacronym{qoe}{QoE}{Quality of Experience}
\newacronym{ml}{ML}{Machine Learning}
\newacronym{ahp}{AHP}{Analytic Hierarchy Process}
\newacronym{lidar}{LIDAR}{Light Detection and Ranging}
\newacronym{sumo}{SUMO}{Simulation of Urban MObility}
\newacronym{wave}{WAVE}{Wireless Access in Vehicular Environment}
\newacronym{c-its}{C-ITS}{Connected Intelligent Transportation System}
\newacronym{dash}{DASH}{Dynamic Adaptive Streaming over HTTP}
\newacronym{http}{HTTP}{HyperText Transfer Protocol}
\newacronym{nt}{NT}{Non-Terrestrial}
\newacronym{ntc}{NTC}{non-terrestrial communication}
\newacronym{ntn}{NTN}{Non-Terrestrial Network}
\newacronym{haps}{HAPS}{High Altitude Platform Station}
\newacronym{hap}{HAP}{High Altitude Platform}
\newacronym{leo}{LEO}{Low Earth Orbit}
\newacronym{meo}{MEO}{Medium Earth Orbit}
\newacronym{geo}{GEO}{Geostationary Earth Orbit}
\newacronym{uav}{UAV}{Unmanned Aerial Vehicle}
\newacronym{nsat}{nSAT}{Nanosatellite}
\newacronym{ehf}{EHF}{extremely high-frequency}
\newacronym{ioe}{IoE}{Internet of Everyone}
\newacronym{gan}{GaN}{Gallium Nitride}
\newacronym{af}{AF}{amplify-and-forward}
\newacronym{csi}{CSI}{channel state information}
\newacronym{ecdf}{ECDF}{empirical cumulative distribution function}
\newacronym{f}{F}{flexible}
\newacronym{fpga}{FPGA}{field programmable gate array}
\newacronym{fov}{FoV}{field-of-view}
\newacronym{km}{KM}{K-means}
\newacronym{kmed}{KMed}{K-medoids}
\newacronym{iab}{IAB}{Integrated Access and Backhaul}
\newacronym{bap}{BAP}{backhaul adaptation protocol}
\newacronym{irs}{IRS}{intelligent reflecting surface}
\newacronym{lsfc}{LSFC}{large-scale fading coefficient}
\newacronym{noma}{NOMA}{non-orthogonal multiple access}
\newacronym{fdm}{FDM}{frequency-division multiplexing}
\newacronym{sdm}{SDM}{space-division multiplexing}
\newacronym{ofdma}{OFDMA}{orthogonal frequency-division multiple access}
\newacronym{oma}{OMA}{orthogonal multiple access}
\newacronym{plos}{pLoS}{probabilistic \ac{los}}
\newacronym{rsma}{RSMA}{rate-splitting multiple access}
\newacronym{scm}{SCM}{spatial channel model}
\newacronym{siso}{SISO}{single input single output}
\newacronym{svd}{SVD}{singular value decomposition}
\newacronym{thz}{THz}{Terahertz}
\newacronym{ula}{ULA}{uniform linear array}
\newacronym{uma}{UMa}{urban macro-cell}
\newacronym{umi}{UMi}{urban micro-cell}
\newacronym{mt}{MT}{mobile terminal}
\newacronym{cu}{CU}{centralized unit}
\newacronym{du}{DU}{distributed unit}
\newacronym{dag}{DAG}{directed acyclic graph}
\newacronym{st}{ST}{spanning tree}
\newacronym{rma}{RMa}{rural macrocell}
\newacronym{inf}{InF}{indoor factory}
\newacronym{ngc}{NGC}{next generation core}
\newacronym{gtp}{GTP}{GPRS Tunnelling Protocol}
\newacronym{tft}{TFT}{Traffic Flow Template}
\newacronym{teid}{TEID}{Tunnel Endpoint Identifier}
\newacronym{tnl}{TNL}{Transport Network Layer}
\newacronym{amf}{AMF}{Access and Mobility Management Function}
\newacronym{ngso}{NGSO}{Non-Geostationary Orbit}
\newacronym{redcap}{RedCap}{Reduced Capability}
\newacronym{ng}{NG}{Next Generation}
\newacronym{fr1}{FR1}{Frequency Range 1}
\newacronym{prach}{PRACH}{Physical Random Access Channel}
\newacronym{ro}{RO}{RACH Occasion}
\newacronym{app}{APP}{Application}
\newacronym{vsat}{VSAT}{Very Small Aperture Terminal}
\newacronym{ack}{ACK}{Acknowledgment}
\newacronym{cwnd}{CWND}{Congestion Window}
\newacronym{rar}{RAR}{Random Access Response}
\newacronym{n6}{N6}{Network 6}
\newacronym{rto}{RTO}{Retransmission Timeout}
\newacronym{ims}{IMS}{IP Multimedia Subsystem}
\newacronym{6gr}{6GR}{6G Radio}
\newacronym{gp}{GP}{Guard Period}
\newacronym{arq}{ARQ}{Automatic Repeat Request}

\title{5G NR Non-Terrestrial Networks: \\ Open Challenges for Full-Stack Protocol Design}

\author{Francesco Rossato, Mattia Figaro, Alessandro Traspadini,~\IEEEmembership{Student Members~IEEE} \\ Takayuki Shimizu, Chinmay Mahabal, Sanjeewa Herath, Chunghan Lee, Dogan Kutay Pekcan, \\ Michele Zorzi,~\IEEEmembership{Fellow Member, IEEE}, Marco~Giordani,~\IEEEmembership{Senior Member, IEEE}
\thanks{F. Rossato, M. Figaro, A. Traspadini, M. Zorzi, and M. Giordani are with the Department of Information Engineering, University of Padova. Padova, Italy. (E-mail: \{francesco.rossato, mattia.figaro, alessandro.traspadini, michele.zorzi, marco.giordani\}@dei.unipd.it).\\
T. Shimizu, C. Mahabal, S. Herath, C. Lee, and D. K. Pekcan are with R\&D InfoTech Labs, Toyota Motor North America Inc., USA. (Email: \{takayuki.shimizu, chinmay.mahabal, sanjeewa.herath, chunghan.lee,dogan.pekcan\}@toyota.com).\\
This work was partially supported by the European Union under the Italian National Recovery and Resilience Plan (NRRP) Mission 4, Component 2, Investment 1.3, CUP C93C22005250001, partnership on ``Telecommunications of the Future'' (PE00000001 -- program ``RESTART''). This work was also partially supported by the European Commission through the European Union’s Horizon Europe Research and Innovation Programme under the Marie Skłodowska-Curie-SE, Grant Agreement No. 101129618, UNITE.}
\vspace{-1em}
}

\newcommand\copyrightnotice{%
\begin{tikzpicture}[remember picture,overlay]
\node[anchor=south,yshift=15pt] at (current page.south) {\fbox{\parbox{\dimexpr\textwidth-\fboxsep-\fboxrule\relax}{
\footnotesize \textcopyright 2026 IEEE. Personal use of this material is permitted.
Permission from IEEE must be obtained for all other uses, in any current or future media,
including reprinting/republishing this material for advertising or promotional purposes,
creating new collective works, for resale or redistribution to servers or lists,
or reuse of any copyrighted component of this work in other works.}}};
\end{tikzpicture}
}

\begin{document}
\maketitle

\copyrightnotice

\begin{abstract}
As \gls{5g} networks continue to evolve, there is a growing interest toward the integration of \glspl{tn} and \glspl{ntn}. Specifically, NTNs leverage space/air base stations such as satellites, \glspl{hap}, and \glspl{uav} for expanding wireless coverage to underserved rural/remote areas, supporting emergency communications, and offloading traffic in highly congested urban environments.
In this paper we focus on the 3GPP 5G NR-NTN standard in the context of satellite communication networks, and highlight critical challenges that must be addressed for proper full-stack protocol design, 
with considerations related to the PHY, MAC, and higher layers. 
We also present simulation results in ns-3 to demonstrate the impact of some of these challenges on the network, as an initial step toward more advanced standardization activities on 3GPP 5G NR-NTN. 
\end{abstract}

\glsresetall


\begin{tikzpicture}[remember picture,overlay]
\node[anchor=north,yshift=-10pt] at (current page.north) {\parbox{\dimexpr\textwidth-\fboxsep-\fboxrule\relax}{
\centering\footnotesize 
This article has been submitted to IEEE for publication. Copyright may change without notice.}};
\end{tikzpicture}

\glsresetall

\section{Introduction}
\label{sec:introduction}
In recent years, there has been a growing interest in the deployment of \glspl{ntn}, which leverage aerial and spaceborne platforms such as satellites, \glspl{hap}, and \glspl{uav} to extend the coverage of \glspl{tn}~\cite{giordani2020non}.
Notably, \gls{leo} satellites offer wide-area connectivity with relatively low latency, and are especially attractive for \glspl{ntn}, as proved by the many commercial Internet access deployments based on LEO constellations.

Satellite communication enables a broad range of applications, such as for connected vehicles, automation, \gls{iot}, aerospace, environmental monitoring, smart grids, and remote management.
Satellites can also provide broadband standalone connectivity in the absence of ground infrastructure \cite{figaro2026experimentalevaluationuavmountedleo}, e.g., in remote or disastered areas.


To enable such diverse applications, the \gls{3gpp} has worked on a new standard to support \gls{ntn} communications, named 5G NR-NTN~\cite{38821}.
However, the integration of NTNs, especially satellites, into the 5G/6G ecosystem presents many challenges compared to \glspl{tn}~\cite{38811}. 
In fact, satellites suffer from more severe path loss and attenuation due to atmospheric factors like scintillation, rain and clouds, as well as Doppler shifts due to orbital mobility \cite{rinaldi2020ntn}. Moreover, the long propagation delay in satellite networks complicates PHY/MAC procedures, especially in terms of channel estimation, resource allocation, scheduling, routing, and handover management. Additionally, satellites create larger coverage areas than terrestrial base stations, and serve a larger number of terminals, which may saturate the available network resources.
Addressing these issues may require substantial modifications to the 5G NR-TN protocol stack, which have not yet been thoroughly analyzed and evaluated by the research community.
In~\cite{giordani2020non}, we took an initial step in this direction by exploring some fundamental challenges for NTN, although the discussion did not reference 3GPP specifications explicitly.
Hosseinian~\emph{et al.}~\cite{hosseinian2021review} reviewed the 5G NTN standardization landscape, although their study was published in 2021, and did not incorporate recent 3GPP advancements beyond Rel. 17.
On the other hand, our recent work~\cite{figaro20255g} presented a more updated overview of the 3GPP 5G NR-NTN standard toward Rel. 20, though we did not analyze current challenges and solutions for protocol~design.

To address these gaps, in this paper we explore open research questions for future 3GPP 5G NR-NTN standardization activities focusing on satellite networks, and formalize possible solutions for proper protocol design. In particular, we review critical aspects related to synchronization and duplexing, resource allocation, retransmissions, coverage, mobility management, routing, and network and transport layers.
Compared to the existing literature, which is largely conceptual with no or limited empirical validation, in this study we run end-to-end system-level simulations in ns-3 to numerically evaluate the effect of these challenges on the 5G NR-NTN protocol stack. We demonstrate the impact of \glspl{gp} in \gls{tdd}, the number of \glspl{harq} processes for retransmissions, the differential delay in large cells, and the \gls{tcp} mechanisms in~NTN.


\section{An Overview of the 3GPP 5G NR-NTN Standard}
\label{sec:ntn_overview}


In this section we review the 3GPP 5G NR-NTN standard (Sec.~\ref{sub:architecture}) and its most relevant specifications (Sec.~\ref{sub:3gpp}), and present a simulation platform for the evaluation of different NTN scenarios (Sec.~\ref{sub:simulation}).


\subsection{General Architecture}
\label{sub:architecture}
The NTN architecture consists of: (i) a \gls{ue}; (ii) an aerial/space station, generally a satellite; and (iii) a ground gateway that interconnects the Radio Access Network (NG-RAN) with the 5G Core (5GC) via the \gls{ng} interface, and with the public Internet via the N6 interface.
The interconnection between the UE and the aerial/space station (gateway) is via a service (feeder) link.

While \gls{geo} satellites have been used for decades for applications such as weather monitoring, positioning, and television broadcasting, \gls{leo} and \gls{meo} constellations have recently gained attention for their ability to deliver low-latency broadband Internet connectivity and flexible coverage.
UAVs, typically operating at low altitudes (a few hundred meters), are flexible and ready-to-use connectivity solutions for temporary events, disaster recovery, and mobile relaying. However, they consume a lot of power for propulsion and hovering, and can stay up in the air only for 30-40 minutes.
HAPs, positioned in the stratosphere (around 20 km), can deliver broad and cost-efficient coverage, though they encounter challenges related to aerodynamics and refueling. 

The 3GPP NTN channel model is specified in~\cite{38811}, building upon a similar model originally developed for TNs~\cite{38901}. It defines several propagation environments for both urban and rural areas, and includes corresponding models for path loss, small-scale fading, and \gls{los} probability. Importantly, it incorporates atmospheric attenuation effects derived from ITU recommendations, especially for tropospheric and ionospheric scintillation.
At the physical layer, satellite stations can be equipped with circular aperture antennas, whereas \glspl{ue} and aerial stations generally mount \gls{upa}, \gls{vsat}, or omni-directional antennas, as described in \cite{38821}.

In terms of carrier frequency, satellites have traditionally operated in legacy bands below 6 GHz to provide wide-area coverage, namely in the L (1–2 GHz) and S (2–4 GHz) bands. Recently, the use of the Ku (12–14 GHz) and Ka (20–30 GHz) bands has been explored, and in some cases adopted, by modern satellite systems, to meet the stringent data rate and low-latency requirements of future wireless services~\cite{giordani2020satellite}.
    
\subsection{The Journey from Rel. 17 to Rel. 20}
\label{sub:3gpp}

Before Rel. 14, the 3GPP focused solely on TNs. 
Rel. 15 remained centered on
5G TNs, but started to explore use cases and requirements for extending cellular services beyond terrestrial boundaries through satellites. 
With Rel. 16, the 3GPP formally launched a new Study Item to adapt the 5G NR physical layer and protocol stack for NTNs. 

Since Rel. 17 (March 2022), the standard has included direct satellite access communication within the first 5G NR-NTN specifications to support applications such as enhanced Mobile Broadband (eMBB) via 5G NR and enhanced Machine-Type Communication (eMTC) via NB-IoT.
The major assumption was that the satellite merely acts as an amplify-and-forward relay between the feeder link and the service link, with no onboard processing capabilities (transparent or bentpipe payload). In this case, the gNB is deployed on the ground, typically co-located with the gateway.
To ensure backward compatibility with 5G TNs and mitigate Doppler effects, nodes are constrained to operating in legacy sub-6 GHz bands (FR1) and using \gls{fdd}. Moreover, Earth-fixed, Quasi-Earth-fixed, and Earth-moving cell configurations are supported to address satellite mobility.

In Rel. 18 (June 2024), the 5G NR-NTN standard was further extended to support new frequency bands in the L and Ka ranges (FR2) to enable higher throughput. Additionally, network functions such as \gls{ue} location verification based on multiple \gls{rtt} measurements with a single satellite and Doppler shift compensation for frequencies higher than 10 GHz have been investigated and developed.
Furthermore, new algorithms have been designed to minimize service disruption during handover due to satellite mobility, as well as new enhancements to \gls{ul} coverage.

In Rel. 19, frozen in December 2025, the 5G NR-NTN standard will support advanced satellite stations with onboard processing capabilities, including signal regeneration, (de)modulation, decoding, and switching/routing. In this sense, the satellite effectively operates as a decode-and-forward gNB (regenerative payload), and the gateway functions as a Transport Network Layer (TNL) node, providing connectivity between the RAN and the 5GC components \cite{lin2025evolution}.
Moreover, Rel. 19 will improve both downlink coverage and uplink capacity through new deployment and multiplexing methods, and specify new mechanisms to signal large service areas for broadcast/multicast services. It will also support \gls{redcap} connectivity in FR1 to enable cost- and energy-efficient \gls{iot} services over satellites.

Finally, Rel. 20 (planned in March 2027) will be a ``bridge'' between the 5G and 6G standards.
It will introduce more GNSS-resilient protocols to handle GNSS outages, jamming, and spoofing. Moreover, Rel. 20 will promote a new radio design for 6G NTN (for waveforms, coding, mobility, multi-connectivity, and deployment scenarios), building upon 5G gaps and limitations.

For a more complete description of the 5G NR-NTN specifications in Rel. 17-20, we refer the readers to our paper~\cite{figaro20255g}.

\begin{table*}[]
    \caption{Relevant 3GPP TDocs for WG1, WG2 and WG3, on the main open challenges for 5G NR-NTN protocol design.} 
    \label{tab:3gpp}
\centering
\footnotesize
\renewcommand{\arraystretch}{0.9} 
\begin{tabular}{|p{2.5cm}|p{3.2cm}|p{9.5cm}|p{1.5cm}|}
\hline
{\color[HTML]{252525} \textbf{Problem}} & {\color[HTML]{252525} \textbf{RAN WG}} & {\color[HTML]{252525} \textbf{Title}}                                                                                            & {\color[HTML]{252525} \textbf{TDoc}}                                                                                                                    \\ \hline

{\color[HTML]{252525} Synchronization} & {\color[HTML]{252525} WG1} & {\color[HTML]{252525}Discussion on NR-NTN GNSS resilient operation using synchronization signals} & {\color[HTML]{252525} R1-2509178} \\ \hline
{\color[HTML]{252525} Synchronization} & {\color[HTML]{252525} WG1} & {\color[HTML]{252525}Positioning, Navigation and Timing (PNT) in 6G NTN-TN harmonization} & {\color[HTML]{252525} R1-2507602} \\ \hline
{\color[HTML]{252525} Synchronization} & {\color[HTML]{252525} WG1} & {\color[HTML]{252525}Random access for NR NTN GNSS resilient operation} & {\color[HTML]{252525} R1-2505695} \\ \hline
{\color[HTML]{252525} Synchronization} & {\color[HTML]{252525} WG1} & {\color[HTML]{252525}Study on GNSS resilient NR-NTN operation} & {\color[HTML]{252525} RP-251933} \\ \hline 
{\color[HTML]{252525} Synchronization} & {\color[HTML]{252525} WG1} & {\color[HTML]{252525}FL Summary \#4 - NR-NTN GNSS resilience} & {\color[HTML]{252525} RP-2506849} \\ \hline 


{\color[HTML]{252525} Resource Allocation}& {\color[HTML]{252525} WG1} & {\color[HTML]{252525} Discussion on DL coverage enhancement for NR-NTN} & {\color[HTML]{252525} R1-2501216} \\ \hline 
{\color[HTML]{252525} Resource Allocation}& {\color[HTML]{252525} WG1} & {\color[HTML]{252525}Discussion on NR-NTN downlink coverage enhancement} & {\color[HTML]{252525} R1-2501114} \\ \hline 
{\color[HTML]{252525} Resource Allocation}& {\color[HTML]{252525} WG1} & {\color[HTML]{252525}Discussion on downlink coverage enhancement for NR-NTN} & {\color[HTML]{252525} R1-2501122} \\ \hline 
{\color[HTML]{252525} Resource Allocation} & {\color[HTML]{252525} WG1} & {\color[HTML]{252525} NR-NTN Uplink Capacity Enhancements} & {\color[HTML]{252525} {R1-2500801}} \\ \hline 
{\color[HTML]{252525} Resource Allocation} & {\color[HTML]{252525} WG1} & {\color[HTML]{252525} FL Summary \#4: NR-NTN downlink coverage enhancements} & {\color[HTML]{252525} {R1-2500042}} \\ \hline 

{\color[HTML]{252525}TDD and FDD} & {\color[HTML]{252525} WG1} & {\color[HTML]{252525} IoT-NTN TDD mode scheduling, timing aspects and channel decoding performance} & {\color[HTML]{252525} R1-2500524} \\ \hline 
{\color[HTML]{252525}TDD and FDD} & {\color[HTML]{252525} WG1} & {\color[HTML]{252525} Discussion on IoT-NTN TDD mode} & {\color[HTML]{252525} R1-2500447} \\ \hline 
{\color[HTML]{252525}TDD and FDD} & {\color[HTML]{252525} WG1} & {\color[HTML]{252525} Discussion on IoT-NTN TDD mode} & {\color[HTML]{252525} R1-2500370} \\ \hline 
{\color[HTML]{252525}TDD and FDD} & {\color[HTML]{252525} WG1} & {\color[HTML]{252525} Discussion on LS Reply on precompensation for NB-IoT NTN TDD mode} & {\color[HTML]{252525} R1-2507086} \\ \hline
{\color[HTML]{252525}TDD and FDD} & {\color[HTML]{252525} WG1} & {\color[HTML]{252525} Discussion on HD-FDD RedCap UEs and eRedCap UEs for FR1-NTN} & {\color[HTML]{252525} R1-2500083} \\ \hline
{\color[HTML]{252525}TDD and FDD} & {\color[HTML]{252525} WG1} & {\color[HTML]{252525} Revised WID for introduction of IoT-NTN TDD mode} & {\color[HTML]{252525} RP-252935} \\ \hline
{\color[HTML]{252525}TDD and FDD} & {\color[HTML]{252525} WG4} & {\color[HTML]{252525} Views on NTN HD-FDD VSAT RF Requirements for Ku-band} & {\color[HTML]{252525} R4-2520822} \\ \hline
{\color[HTML]{252525}TDD and FDD} & {\color[HTML]{252525} WG4} & {\color[HTML]{252525} Discussion on NTN enhancement HD-FDD for Ku band small-type Antenna} & {\color[HTML]{252525} R4-2521501} \\ \hline


{\color[HTML]{252525}HARQ} & {\color[HTML]{252525} WG1} & {\color[HTML]{252525} Maintenance on disabling of HARQ feedback for IoT NTN} & {\color[HTML]{252525} R1-2400470} \\ \hline 
{\color[HTML]{252525}HARQ} & {\color[HTML]{252525} WG2} & {\color[HTML]{252525} Views on 6G User Plane: HARQ and Scheduling} & {\color[HTML]{252525} R2-2508383} \\ \hline 
{\color[HTML]{252525}HARQ} & {\color[HTML]{252525} WG2} & {\color[HTML]{252525} FLS\#1 on disabling of HARQ feedback for IoT NTN} & {\color[HTML]{252525} R1-2401497} \\ \hline 
{\color[HTML]{252525}HARQ} & {\color[HTML]{252525} WG2} & {\color[HTML]{252525} Report from Break-Out Session on NR NTN and IoT NTN} & {\color[HTML]{252525} R2-2401543} \\ \hline 
{\color[HTML]{252525}HARQ} & {\color[HTML]{252525} WG2} & {\color[HTML]{252525} Summary of [AT121bis-e][103][IoT NTN Enh] HARQ enhancements} & {\color[HTML]{252525} R2-2304243} \\ \hline 
{\color[HTML]{252525}HARQ} & {\color[HTML]{252525} WG2} & {\color[HTML]{252525} Detection of consecutive HARQ feedback failures in NB-IoT NTN} & {\color[HTML]{252525} R2-2500462} \\ \hline 

{\color[HTML]{252525} Handover and Paging} & {\color[HTML]{252525} WG2} & {\color[HTML]{252525} New WID: E-UTRA TN to NR NTN handover enhancements} & {\color[HTML]{252525} {RP-251878}} \\ \hline 
{\color[HTML]{252525} Handover and Paging} & {\color[HTML]{252525} WG2} & {\color[HTML]{252525} Revise WID on E-UTRA TN to NR NTN handover enhancements} & {\color[HTML]{252525} {RP-252890}} \\ \hline 
{\color[HTML]{252525} Handover and Paging} & {\color[HTML]{252525} WG3} & {\color[HTML]{252525} Discussion on inter-gNB RACH-less HO in NTN} & {\color[HTML]{252525} {R3-250217}} \\ \hline

\end{tabular}
\end{table*}

\subsection{A New Simulation Platform for NTN Research}
\label{sub:simulation}

While the current state of the art has established a robust theoretical framework for \gls{ntn}, particularly regarding signal propagation and physical layer dynamics, our work attempts to advance the field by bridging the critical gap toward more practical implementation. 
Notably, experimental and simulation activities are usually monopolized by a small set of well-funded  companies and system operators that possess the resources to build prototyping platforms and/or simulators in-house.
Yet, software and hardware architectures are generally closed-source, and provide little user customization, which prevents accurate verification of the results and simulation assumptions that may have been introduced. 

To fill these gaps, we recently released \texttt{ns3-NTN}, an open-source ns-3 module to simulate satellite communication networks~\cite{sandri23implementation}.
Unlike traditional link-level simulators, \texttt{ns3-NTN} supports full-stack end-to-end simulations based on the most recent 3GPP 5G NR-NTN specifications. 
It implements several key features, including the 3GPP NTN path-loss, channel, absorption, and antenna models based on~\cite{38811}, an Earth-Centered, Earth-Fixed (ECEF) coordinate system, and  accurate satellite-specific propagation delay models, timing advance mechanisms~\cite{38214} and adjusted \gls{rrc} and \gls{harq} timers~\cite{38821}.
The module has been validated in~\cite{sandri23implementation,figaro20255g} against 3GPP calibration results, and is emerging as one of the most accurate and accessible tools for NTN simulations.

In light of this, in Sec.~\ref{sec:challenges} we will use \texttt{ns3-NTN} to evaluate the performance of several 5G NR-NTN protocol implementations in representative NTN~scenarios.

\section{Open Challenges for \\ 5G NR-NTN Protocol Design}
\label{sec:challenges}

 In this section, we present and evaluate the current research challenges for proper 3GPP 5G NR-NTN protocol design, based on ns-3 simulations. Specifically, we focus on key topics including synchronization, resource allocation, duplexing, \gls{harq}, mobility management, and network and transport layers.  Moreover, we discuss potential protocol solutions based on 3GPP ongoing efforts and research activities, as summarized in Table~\ref{tab:3gpp}.

\subsection{Synchronization}
\label{sub:sync}
Time and frequency synchronization is essential for reliable network connectivity and   accurate decoding.
To do so, in 5G \gls{nr}, \glspl{ue} detect \glspl{pss} and \glspl{sss} before network connection.
However, in NTN, the synchronization process is more challenging.

In the frequency domain, synchronization requires estimating and correcting frequency offsets due to Doppler imperfections.
In NTN, satellites (except GEO) move fast with respect to Earth, and introduce strong Doppler shift variations over time.
For instance, a \gls{leo} satellite at an altitude of 600~Km leads to a maximum Doppler shift of around 48 (480) ~kHz in the S (Ka) band, vs. around 500–700 Hz in \glspl{tn}~\cite{38811}.
Since \gls{3gpp} Rel. 17, Doppler shifts can be pre-compensated for UL traffic based on the GNSS \gls{ue} position within the cell, which determines the relative geometry between the UE and the satellite, and the satellite ephemeris, which provides precise information about the satellite's orbit and velocity.

To prevent inter-slot interference and overlap, \gls{ul} frames transmitted by the UEs must be aligned with the corresponding DL frames at the \gls{gnb}.
In \glspl{tn}, a \gls{ta} mechanism was introduced to compensate for possible propagation delays in the cell, and handle small time offsets in UL and DL.
However, NTN requires a much longer \gls{ta} than that supported in \glspl{tn} (e.g., up to 10 ms for GEO satellites) due to the larger size of satellite cells.
To address this issue, \gls{3gpp} Rel. 17 introduced an open-loop hybrid \gls{ta} mechanism for NTN, which consists of two components.
The first component is a common \gls{ta} derived from the network parameters and equal for all the \glspl{ue} within the cell. The second component is a \gls{ue}-specific \gls{ta}, based on the individual \gls{ue} position within the cell. The standard assumes that each \gls{ue} is equipped with a functional GNSS receiver to autonomously calculate this \gls{ta}. 


\begin{figure}[t!]
    \centering
            \definecolor{8proc}{RGB}{112,105,157}
\definecolor{16proc}{RGB}{197,105,161}
\definecolor{32proc}{RGB}{255,128,125}
\definecolor{64proc}{RGB}{255,185,56}
\definecolor{off}{RGB}{255,127,14}

\definecolor{8proc_edge}{RGB}{164, 156, 214}
\definecolor{16proc_edge}{RGB}{232, 151, 200}
\definecolor{32proc_edge}{RGB}{255, 150, 148}
\definecolor{64proc_edge}{RGB}{250, 200, 107}
\definecolor{off_edge}{RGB}{255, 157, 71}

\usetikzlibrary{positioning}

\begin{tikzpicture}

\begin{axis}[
    name=thr,
    ybar=0pt, 
    width=8cm,
    bar width=20pt,
    enlarge x limits=0.2, 
    ylabel={Cell throughput [Mbps]},
    xtick={1, 2, 3},
    xticklabels={
      \shortstack{LEO 600~km},
      \shortstack{MEO 1200~km},
      \shortstack{GEO 36\,000~km}
    },
    ymin=0, ymax=110,
    ymajorgrids=true,
    grid style=dashed,
    tick label style={font=\footnotesize},
    ylabel style={font=\footnotesize},
    xlabel style={font=\footnotesize}
]

\addplot[fill=8proc, draw=black] coordinates {(1, 99) (2, 100) (3, 97)};
\addplot[fill=16proc, draw=black,postaction={pattern=north east lines}] coordinates {(1, 16) (2, 11) (3, 7)};
\end{axis}

\node[
    yshift=160pt,
    xshift=80pt,
    anchor=south, 
    draw, 
    fill=white, 
    inner sep=3pt, 
    font=\small, 
] (customlegend) {
    {
    \setlength{\tabcolsep}{2pt}
    \begin{tabular}{c l @{\hspace{12pt}} c l}
    \tikz \draw[fill=8proc, draw=black] (0,0) rectangle (0.3,0.3); & Cell-center UEs &
    \tikz \draw[fill=16proc, draw=black,postaction={pattern=north east lines}] (0,0) rectangle (0.3,0.3); & Sparse UEs
    \end{tabular}
    }
};

\end{tikzpicture}
        \caption{Impact of the differential delay among different UEs on the throughput. We consider $N_u=4$ \glspl{ue} communicating in S band and generating UL UDP data with a source rate of 25 Mbps each, so the cumulative cell data rate is 100 Mbps. We compare the case in which all UEs are deployed around the cell center (plain bars) vs. the case in which they are uniformly distributed across the cell (striped~bars).}
        \label{fig:diffdel_plot}
        \vspace{-0.5em}
\end{figure}
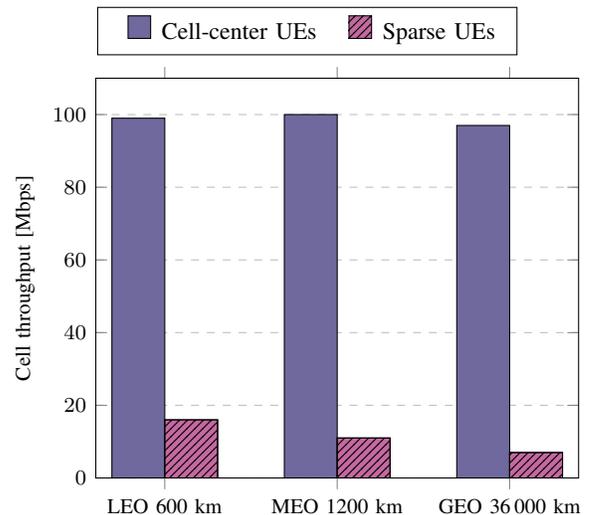

In this paper, we study the case in which a GNSS reference measure is unavailable, for example in urban canyons or indoors, or due to GNSS interference or spoofing. Under this condition, \glspl{ue} cannot compute the required Doppler pre-compensation or \gls{ta} parameters, resulting in frequency and time synchronization errors, respectively.
To demonstrate this effect, in Fig. \ref{fig:diffdel_plot} we use our \texttt{ns3-NTN} module to plot the cumulative UL throughput of a satellite cell in which $N_u=4$ UEs are clustered around the cell center (``Cell-center UEs''), or uniformly distributed across the whole cell (``Sparse UEs'').
We focused on S band, and configured the link budget so that all \glspl{ue} experience a sufficient \gls{snr} to maintain the target source rate (25 Mbps), regardless of their position in the cell. 
In the first scenario the cell throughput is close to the cell source rate (100 Mbps). This is because all UEs are co-located, and experience nearly identical propagation delays, therefore 
the network can effectively compensate for (mostly minor) synchronization offsets through the baseline 5G \gls{nr} \gls{ta} or guard periods. In contrast, in the second scenario, the throughput deteriorates by more than $80\%$, especially when the satellite altitude, and so the cell size, increase.
In this case, the differential delay between UEs at the cell center and those at the cell edge may introduce scheduling errors between the time data is expected to arrive at the gNB and when it is actually received. To solve this, the network would require longer TA values than those supported in 5G NR, leading to throughput degradation. 
Ongoing research, such as in~\cite{lin2021dopplershiftestimation5g}, explores alternative solutions for UEs without GNSS capabilities, such as low-complexity \gls{iot} devices, as recommended for 3GPP Rel. 18+. 

\subsection{Resource Allocation}



Resource allocation, especially in the \gls{ul}, is currently based on a combination of scheduling requests and grants to and from the gNB and the UEs, respectively, which may take up to two \glspl{rtt}. While this delay is tolerable in \glspl{tn}, it may become prohibitively large in NTNs.
Therefore, the \gls{3gpp} recommends minimizing the number of signals to be exchanged to reduce latency and overhead \cite{38821}.

An NTN-gNB serves a larger number of UEs than in the terrestrial case, which may saturate the available resources and complicate channel access.
For example, for a single LEO satellite at an altitude of 600 km with a beamwidth of 40\textdegree, the approximate coverage area is around 1 million km$^2$ (vs. 80 km$^2$ for a 5G \gls{nr} gNB) and, with an average density of 100 UEs/km$^2$ for pedestrian UEs \cite[Table B.2-1]{38821}, it would amount to up to 100 million \glspl{ue} (vs. around 8000 for a 5G \gls{nr} gNB). To partially solve this issue, satellites operate through spot beams to partition coverage into smaller geographic areas, and allocate resources to a limited set of UEs via \gls{sdma}.


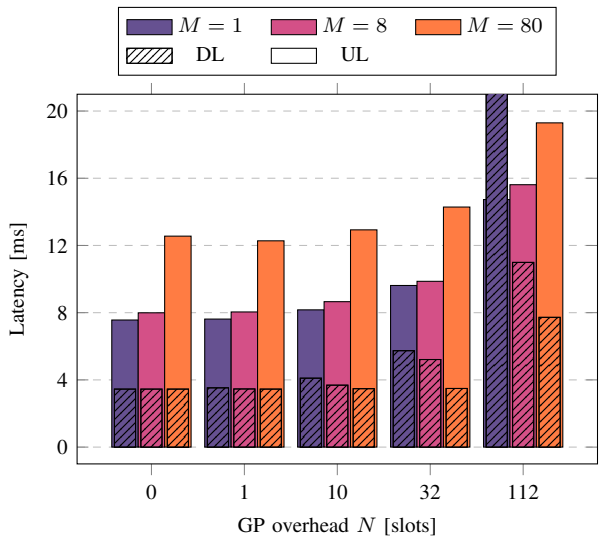
\begin{figure}
    \centering
    \begin{tikzpicture}

\definecolor{kaBlue}{RGB}{0,114,189}
\definecolor{kaRed}{RGB}{217,83,25}
\definecolor{sLightBlue}{RGB}{173,216,230}
\definecolor{sLightRed}{RGB}{255,182,193}

\definecolor{DarkBlue}{HTML}{003f5c}
\definecolor{LightBlue}{HTML}{2f4b7c}
\definecolor{DarkPurple}{HTML}{665191}
\definecolor{LightPurple}{HTML}{a05195}
\definecolor{DarkPink}{HTML}{d45087}
\definecolor{LightPink}{HTML}{f95d6a}
\definecolor{DarkOrange}{HTML}{ff7c43}
\definecolor{LightOrange}{HTML}{ffa600}

\begin{axis}[
    width=8.5cm, 
    height=6.5cm,
    ybar,
    bar width=15pt,
    ylabel={Latency [ms]},
    xlabel={GP overhead $N$ [slots]},
    enlarge x limits=0.2,
    symbolic x coords={0, 1, 10, 32, 112},
    xticklabels={0, 1, 10, 32, 112},
    xtick=data,
    xticklabel style={align=center},
    ymin=-1, ymax=21,
    ymajorgrids=true,
    ytick={0, 4, 8, 12, 16, 20},
    yticklabels={0, 4, 8, 12, 16, 20},
    grid style=dashed,
    tick label style={font=\footnotesize},
    ylabel style={font=\footnotesize},
    xlabel style={font=\footnotesize},
    legend style={font=\footnotesize},
    legend style={at={(0.5,1.05)}, anchor=south, legend columns=3,/tikz/every even column/.append style={column sep=0.3cm}}, 
    nodes near coords={},
    nodes near coords style={font=\footnotesize, yshift=6pt},
    every node near coord/.style={}
]

\addplot+[ybar, bar shift=-10pt, area legend, fill=DarkPurple, draw=black, bar width=-10pt] coordinates { 
(0, 7.563959)
(1, 7.619854)
(10, 8.1677645)
(32, 9.618673)
(112, 14.731076)
};
\addlegendentry{$M = 1$}

\addplot+[ybar, bar shift=0pt, area legend, fill=DarkPink, draw=black, bar width=10pt] coordinates {
(0, 7.9923975)
(1, 8.042887)
(10, 8.6547375)
(32, 9.8664815)
(112, 15.615716)
};
\addlegendentry{$M = 8$}

\addplot+[ybar, bar shift=10pt, area legend, fill=DarkOrange, draw=black, bar width=10pt] coordinates { 
(0, 12.5548995)
(1, 12.2763175)
(10, 12.9287475)
(32, 14.286306)
(112, 19.295817)
};
\addlegendentry{$M = 80$}

\addplot+[ybar, bar width=8pt, area legend, bar shift=10pt, fill=white, draw=black, postaction={pattern=north east lines}] coordinates {
(0, 0)};
\addlegendentry{DL}

\addplot+[ybar, bar width=8pt, area legend, bar shift=10pt, fill=white, draw=black] coordinates {
(0, 0)};
\addlegendentry{UL}

\addplot+[ybar, bar shift=-10pt, fill=DarkPurple, draw=black, bar width=8pt, postaction={pattern=north east lines}] coordinates {
(0, 3.451783)
(1, 3.526783)
(10, 4.101783)
(32, 5.737495)
(112, 1517.791063)
};

\addplot+[ybar, bar width=8pt, bar shift=0pt, draw=black,
    fill=DarkPink, postaction={pattern=north east lines}] coordinates {
(0, 3.451783)
(1, 3.460711)
(10, 3.685711)
(32, 5.212495)
(112, 10.991063)
};

\addplot+[ybar, bar width=8pt, bar shift=10pt, fill=DarkOrange, draw=black, postaction={pattern=north east lines}] coordinates {
(0, 3.451783)
(1, 3.451783)
(10, 3.476783)
(32, 3.489279)
(112, 7.721423)
};

\end{axis}
\end{tikzpicture}
    \caption{Application latency as a function of $M$ (the number of consecutive DL slots) and $N$ (the number of GP slots where no transmissions can be scheduled). We consider a DL UDP flow with a source rate of 10 Mbps, and an uplink feedback flow with a source rate of 5 Kbps, between a terrestrial UE and a LEO satellite at an altitude of 600 km operating in the Ka band with numerology 3.}
    \label{fig:gp_thr}
    \vspace{-0.5em}
\end{figure}

\subsection{Time/Frequency Division Duplexing}
Most satellite networks, as well as 4G/LTE cellular networks and previous generations, are traditionally \gls{fdd}-based, where \glspl{ue} and satellites transmit and receive on separate frequency bands for UL and DL. This setup enables full-duplex communication, avoids the need for strict time synchronization, and facilitates interference mitigation as UL and DL are on orthogonal frequencies.
In contrast, 5G \gls{nr} networks are predominantly \gls{tdd}-based, which offers several benefits such as channel reciprocity, reduced hardware complexity, and a more efficient spectrum usage.
However, this approach requires a \gls{gp} between consecutive UL and DL slots to prevent interference. The duration of this \gls{gp} must be proportional to the distance, and so the propagation delay, between the UE and the satellite-gNB, which is particularly large in NTN.
According to the \gls{3gpp}~\cite{38811}, assuming numerology 0, it ranges between 4 and 14 ms for a \gls{leo} satellite at 600 km, and between 480 and 540 ms for a \gls{geo} satellite. For comparison, the duration of a 14-OFDM-symbol slot is 1 ms, which translates into a significant TDD overhead and throughput degradation.
For the sake of channel reciprocity, hardware complexity, and frequency diversity, the \gls{3gpp} considers the use of \gls{tdd} only for \glspl{hap} or \gls{leo} satellites~\cite{38811}.  

To demonstrate this issue, we use our \texttt{ns3-NTN} simulator (see Sec.~\ref{sub:simulation}) to evaluate the impact of the \gls{gp} on the network latency, and the results are reported in Fig. \ref{fig:gp_thr}. 
We consider a DL traffic application in the Ka band with numerology 3, and assume the network can aggregate $M=\{1,8,80\}$ consecutive slots for \gls{dl} data, corresponding to a single slot, a subframe, and an entire frame, respectively. 
The GP duration is equal to $N\in\{0,1,10,31,112\}$ slots, corresponding to a duration of $\{0 \text{ (ideal)},0.125,1.25,4,14\}$ ms, respectively, as expected for LEO links.
\Gls{ul} traffic mainly consists of small feedback or channel state packets, and always occupies one slot.
We observe that the impact of the GP is not negligible.
For $M=80$, the DL (UL) latency increases from 3 to 8 ms (12 to 19 ms) when $N$ grows from 0 to 112, given that no transmissions can be scheduled during \glspl{gp}.
This effect is even more pronounced when $M$ is small:
in this case, the system switches between DL and UL more frequently, which requires more GPs, thereby increasing the overall overhead and latency.
Moreover, the DL latency decreases as $M$ increases, as aggregating more DL slots provides more frequent DL transmission opportunities, though at the expense of the UL~latency, making the network less responsive. This behavior is more evident as $N$ increases, due to the more severe GP overhead in the frame~structure.  

 To partially address this issue, we recently proposed a novel TDD slot allocation mechanism that permits to schedule additional transmissions during \glspl{gp} to reduce the overhead, as long as they do not interfere with other concurrent transmissions~\cite{traspadini2024time}. Simulation results showed that this approach can improve the network capacity compared to a baseline scheme that leaves \glspl{gp} unused.


\subsection{HARQ}
\gls{harq} requires the transmitter to wait for an ACK or NACK from the receiver before scheduling new transmissions or retransmissions, respectively, to improve reliability. 
However, the long RTT in NTN causes the HARQ process to stall while waiting for ACKs, which increases the communication delay.
Specifically, the RTT normally exceeds the maximum duration of 5G \gls{nr} HARQ timers (which defines how long to wait for an ACK) and the maximum number of HARQ processes $n$ (which defines how many HARQ transmissions can be handled in parallel while waiting for ACKs).


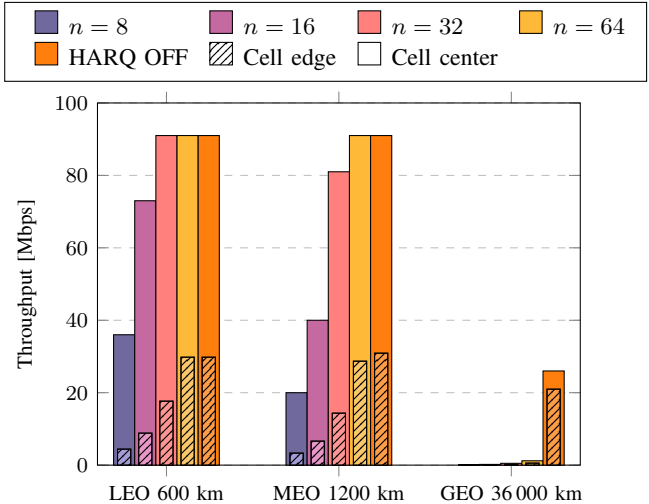
\begin{figure}[t!]
    \centering
            \definecolor{8proc}{RGB}{112,105,157}
\definecolor{16proc}{RGB}{197,105,161}
\definecolor{32proc}{RGB}{255,128,125}
\definecolor{64proc}{RGB}{255,185,56}
\definecolor{off}{RGB}{255,127,14}

\definecolor{8proc_edge}{RGB}{164, 156, 214}
\definecolor{16proc_edge}{RGB}{232, 151, 200}
\definecolor{32proc_edge}{RGB}{255, 150, 148}
\definecolor{64proc_edge}{RGB}{250, 200, 107}
\definecolor{off_edge}{RGB}{255, 157, 71}

\usetikzlibrary{positioning}

\begin{tikzpicture}

\begin{axis}[
    width=8cm, 
    height=6.4cm,
    name=data,
    ybar=0pt, 
    bar width=8pt,
    enlarge x limits=0.2, 
    ylabel={Throughput [Mbps]},
    xtick={1, 2, 3},
    xticklabels={
      \shortstack{LEO 600~km},
      \shortstack{MEO 1200~km},
      \shortstack{GEO 36\,000~km}
    },
    ymin=0, ymax=100,
    ymajorgrids=true,
    grid style=dashed,
    tick label style={font=\footnotesize},
    ylabel style={font=\footnotesize},
    xlabel style={font=\footnotesize}
]

\addplot[fill=8proc, draw=black] coordinates {(1, 36) (2, 20) (3, 0.1)};
\addplot[fill=16proc, draw=black] coordinates {(1, 73) (2, 40) (3, 0.2)};
\addplot[fill=32proc, draw=black] coordinates {(1, 91) (2, 81) (3, 0.5)};
\addplot[fill=64proc, draw=black] coordinates {(1, 91) (2, 91) (3, 1.2)};
\addplot[fill=off, draw=black] coordinates {(1, 91) (2, 91) (3, 26)};
\end{axis}

\begin{axis}[
    at={(data.south west)},
    anchor=south west,
    width=8cm,
    ybar=0pt, 
    bar width=8pt,
    enlarge x limits=0.2, 
    ylabel={Throughput [Mbps]},
    xtick={1, 2, 3},
    xticklabels={
      \shortstack{LEO 600~km},
      \shortstack{MEO 1200~km},
      \shortstack{GEO 36\,000~km}
    },
    ymin=0, ymax=100,
    ymajorgrids=true,
    grid style=dashed,
    tick label style={font=\footnotesize},
    ylabel style={font=\footnotesize},
    xlabel style={font=\footnotesize},
    hide axis]
\addplot[fill=8proc_edge, bar width=5pt, bar shift=-16pt, draw=black, postaction={
        pattern=north east lines
    }] coordinates {(1, 4) (2, 3) (3, 0)};
\addplot[fill=16proc_edge, bar width=5pt, bar shift=-8pt, draw=black, postaction={
        pattern=north east lines
    }] coordinates {(1, 8) (2, 6) (3, 0.1)};
\addplot[fill=32proc_edge, bar width=5pt, bar shift=0pt, draw=black, postaction={
        pattern=north east lines
    }] coordinates {(1, 16) (2, 13) (3, 0.2)};
\addplot[fill=64proc_edge, bar width=5pt, bar shift=8pt, draw=black, postaction={
        pattern=north east lines
    }] coordinates {(1, 27) (2, 26) (3, 0.5)};
\addplot[fill=off_edge, bar width=5pt, bar shift=16pt, draw=black, postaction={
        pattern=north east lines
    }] coordinates {(1, 27) (2, 28) (3, 19)};
\end{axis}

\node[
    at={(data.north)},
    draw, 
    fill=white, 
    inner sep=3pt, 
    font=\small, 
    anchor=south, 
    yshift=8pt, 
    xshift=-5pt
] (customlegend) {
    {
    \setlength{\tabcolsep}{2pt}
    \begin{tabular}{
    l l @{\hspace{8pt}}
    l l @{\hspace{8pt}}
    l l @{\hspace{8pt}}
    l l
    }
    \tikz \draw[fill=8proc, draw=black] (0,0) rectangle (0.3,0.3); & $n=8$ &
    \tikz \draw[fill=16proc, draw=black] (0,0) rectangle (0.3,0.3); & $n=16$ &
    \tikz \draw[fill=32proc, draw=black] (0,0) rectangle (0.3,0.3); & $n=32$ & 
    \tikz \draw[fill=64proc, draw=black] (0,0) rectangle (0.3,0.3); & $n=64$  \\ 
    \tikz \draw[fill=off, draw=black] (0,0) rectangle (0.3,0.3); & HARQ OFF &
    \tikz \draw[fill=white, pattern=north east lines] (0,0) rectangle (0.3,0.3); & Cell edge &
    \tikz \draw[fill=white] (0,0) rectangle (0.3,0.3); & Cell center
    \end{tabular}
    }
};

\end{tikzpicture}
        \caption{Application throughput in the S band (at the cell center and at the cell edge) as a function of the satellite altitude, for different HARQ configurations. We change the number of HARQ processes, $n$, vs. a benchmark scheme where HARQ is disabled. We consider $N_u = 4$ UEs generating UL UDP data with a source rate of 25 Mbps, so the cumulative cell data rate is 100 Mbps. }
        \label{fig:harq_plot}
\vspace{-0.5em}
\end{figure}

One option is to increase the number of HARQ processes beyond $16$ (the current limit for 5G NR \glspl{tn}) based on the \gls{rtt} of the network. While the 3GPP suggests to configure $32$ processes for LEO satellites, MEO and GEO satellites, as well as cell-edge cases, may require even more to avoid the risk of HARQ stall, due to the increased propagation delay. Moreover, for \gls{geo} satellites, \gls{harq} feedback can be disabled in the presence of ARQ retransmissions at the \gls{rlc} layer, according to Rel. 17 \gls{nr}-\gls{ntn}.

In Fig.~\ref{fig:harq_plot}, we evaluate via \texttt{ns3-NTN} simulations the throughput in the S band for different satellite orbits as a function of the number of per-\gls{ue} \gls{harq} processes, $n$.
The throughput is maximized when HARQ is disabled (HARQ OFF), although with negative implications in terms of reliability, especially in low \gls{snr} regimes.
While retransmissions
can be delegated to the higher layers, this approach introduces
additional latency and protocol overhead.
We observe that, as the satellite altitude increases, the system needs to increase $n$ to compensate for the longer propagation delay, and prevent HARQ from stalling.
At the cell center, for a LEO satellite at 600 km, the maximum throughput is 91 Mbps, which is obtained with $n=32$ (the 3GPP 5G NR-NTN reference value), while at 1200 km it is $n=64$. Eventually, we reach a saturation point where \gls{harq} is no longer the bottleneck. 
Our simulations also indicate that \gls{harq} is impractical for \gls{geo} satellites, which would require up to $n=600$ processes to fully utilize the radio link~\cite[Table 7.3.3.1.1-1]{38811}. However, this approach would increase the complexity at both the \gls{ue} and the gNB.
Finally, in Fig.~\ref{fig:harq_plot} we also plot the throughput at the cell edge. Although the overall trend is similar than at the cell center, the maximum throughput is substantially lower ($-70\%$) due to the longer link and more severe path loss.



\subsection{Handover and Paging}
In 5G \gls{nr} \glspl{tn}, UEs are generally mobile, while the gNB is assumed to be stationary. In NTNs, instead, both communication endpoints may be subject to mobility. 
For example, a LEO satellite at an altitude of 600 km moves rapidly at speeds of approximately $7-8 $ km/s~\cite{leo-sat}, so the orbital period is 96.7 min, and the visibility period over a given ground footprint is only 7 to 10 min. Notably, mobility affects the whole protocol stack, particularly paging and handover.

Paging requires the Access and Mobility Management Function (AMF) in the 5GC to locate a UE in idle mode within a Registration Area to deliver incoming data.
However, unlike in \glspl{tn},
in NTNs there is no one-to-one correspondence between a satellite-gNB and a specific Registration Area, due to the continuous movement of the~satellite. 
Therefore, it is often unclear whether UEs remain in the same Registration Area, which can lead to missed paging occasions for the UEs. This situation is further complicated by the fact that UEs themselves may also~move. 

Handover is the process of transferring an active \gls{ue} connection from one gNB to another. 
Due to the large coverage areas and high-speed mobility of a satellite-gNB,
a potentially very large number of ground UEs may need to perform handovers simultaneously, resulting in significant signaling overhead at the \gls{ran} and service continuity challenges. For example, the 3GPP estimates that, for a cell diameter of 1000 km, and assuming around 65\,000 UEs (i.e., the average device density is 0.08 UEs/km$^2$), the average time required for handover is as large as 132 s~\cite{38821}. 
Moreover, while in \glspl{tn} adjacent gNBs are generally interconnected via fiber links (i.e., X2 interface), handover coordination in satellite networks relies on wireless ISLs based on optical or THz connections, which are more susceptible to bottlenecks and~congestion.
      
To overcome these issues, within Rel. 17, the 3GPP has explored a cell-fixed handover procedure in which the cell footprint is anchored to a fixed point on Earth through dynamic beam steering, rather than to the satellite. In this way, handover occurs between satellites rather than between moving cells. Compared to the classical cell-moving approach, the cell-fixed model substantially reduces the handover frequency, and provides more stable \gls{qos} over time.
Another approach is to use the ephemeris data of LEO satellites to determine their footprint locations and velocity over time. 
As such, the network can estimate the satellites' serving regions on Earth at any given time, and trigger timely and optimized handover accordingly.

\begin{figure}[t!]
    \centering
    \begin{minipage}[b]{0.45\textwidth}
        \centering
        \begin{tikzpicture}

\definecolor{udpColor}{RGB}{217,83,25} 
\definecolor{tcpColor}{RGB}{0,114,189} 

\begin{axis}[
    ybar,
    bar width=12pt,
    symbolic x coords={0,10,15,20,30},
    xticklabels={600,10000,15000,20000,30000},
    xtick=data,
    ymin=0, ymax=11,
    ylabel={Throughput [Mbps]},
    xticklabel style={align=center},
   xticklabels={
      \shortstack{LEO \\ 600~km},
      \shortstack{MEO \\ 10\,000~km},
      \shortstack{MEO \\ 15\,000~km},
      \shortstack{MEO \\ 20\,000~km},
      \shortstack{GEO \\ 36\,000~km}
    },
    ymajorgrids=true,
    grid style=dashed,
    tick label style={font=\footnotesize},
    ylabel style={font=\footnotesize},
    xlabel style={font=\footnotesize},
    area legend,
    legend style={font=\footnotesize, at={(0.5,1)}, legend columns=2, anchor=north, column sep=1.5em},
]

\addplot+[ybar, bar shift=0pt, fill=udpColor, draw=black] coordinates {
    (0,10)
    (10,9.99584)
    (15,9.89072)
    (20,9.972)
    (30,9.99336)
};

\addplot+[ybar, bar width=6pt, bar shift=0pt, fill=tcpColor, draw=black] coordinates {
    (0,10)
    (10,10)
    (15,3.8)
    (20,1.2)
    (30,0.56)
};

\legend{UDP, TCP}

\end{axis}

\end{tikzpicture}
    \end{minipage}\\\vspace{0.5cm}
    \begin{minipage}[b]{0.45\textwidth} 
        \centering
        \hspace*{-0.8cm}
        \begin{tikzpicture}

\definecolor{udpColor}{RGB}{217,83,25} 
\definecolor{tcpColor}{RGB}{0,114,189} 

\begin{axis}[
    ybar,
    bar width=12pt,
    symbolic x coords={0,10,15,20,30},
        xticklabels={0,10000,15000,20000,30000},
    xtick=data,
    ymin=0, ymax=1500,
    ylabel={Latency [ms]},
    xticklabel style={align=center},
    xticklabels={
      \shortstack{LEO \\ 600~km},
      \shortstack{MEO \\ 10\,000~km},
      \shortstack{MEO \\ 15\,000~km},
      \shortstack{MEO \\ 20\,000~km},
      \shortstack{GEO \\ 36\,000~km}
    },
    ymajorgrids=true,
    grid style=dashed,
    tick label style={font=\footnotesize},
    ylabel style={font=\footnotesize},
    xlabel style={font=\footnotesize},
    area legend,
    legend style={font=\footnotesize, at={(0.5,1)}, legend columns=2, anchor=north, column sep=1.5em},
]

\addplot+[ybar, bar shift=0pt, fill=udpColor, draw=black] coordinates {
    (0,1.442855)
    (10,34.442855)
    (15,51.467855)
    (20,67.442855)
    (30,101.467855)
};

\addplot+[ybar, bar width=6pt, bar shift=0pt, fill=tcpColor, draw=black] coordinates {
    (0,1.442855)
    (10,454.041063)
    (15,890.141063)
    (20,1188.428563)
    (30,1263.491063)
};

\legend{UDP, TCP}

\end{axis}

\end{tikzpicture}
    \end{minipage}\hfill
    \caption{Average DL throughput (top) and latency (bottom) with UDP and TCP Cubic vs. the altitude of the satellite, $h$. The application generates data at a source rate of 10 Mbps, and both the transmitter (satellite-gNB) and the receiver (UE) are equipped with \gls{vsat} antennas to operate in good SNR regimes. The operating frequency is in the Ka band.} 
    \label{fig:tcp_udp_performance}
    \vspace{-0.5em}
\end{figure}
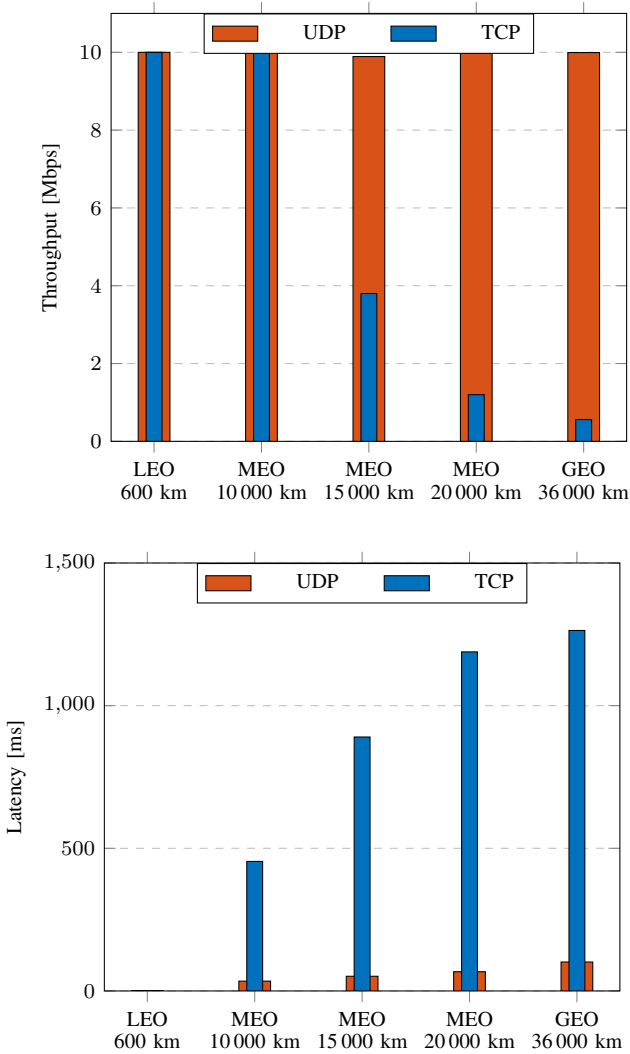

\subsection{Routing}
In the satellite scenario, UEs operate at long distance, and therefore experience large delays. 
At the network level, routing via ISL for regenerative payloads faces several critical challenges. In particular, routing tables may quickly become obsolete (especially when the topology changes due to mobility or link failure, e.g., for LEO satellites), and require frequent and periodic updates, thus increasing the communication overhead.
Moreover, LEO constellations may consist of thousands of satellites, resulting in an extremely large and dynamic routing domain compared to conventional routing architectures.
Additionally, routing protocols must be designed to protect against security threats, for example via  
a trusted central authority, to authenticate nodes and prevent accidental or illegitimate destruction, loss, alteration, or unauthorized access to routing tables.

The 5G NR-NTN standard suggests several mechanisms to handle the unique challenges of satellite links. For example, the predictability of satellite orbits permits to create a new class of proactive routing protocols able to compute pre-established optimal routing paths  based on known statistics.


\subsection{Transport Layer}
According to the \gls{3gpp}, the 5GC is agnostic to whether the radio access is terrestrial or non-terrestrial. As a result, the transport layer shall remain standard, with no modifications for NTN. 
Therefore, we used our \texttt{ns3-NTN} module to evaluate the impact of different transport layer configurations on the \gls{ran}. 
In Fig.~\ref{fig:tcp_udp_performance} we focus on the Ka band, and illustrate the average throughput of a \gls{tcp} Cubic vs. \gls{udp} connection as a function of the altitude $h$ of the satellite. 
We observe that \gls{udp} is not significantly affected by the altitude of the satellite, and the throughput consistently approaches the source rate of the application, set to 10 Mbps. 
This is because UDP does not require time-consuming handshake or ACK mechanisms, so data transmission can progress at full capacity, regardless of the underlying propagation delay. Its lightweight design, simplicity, low latency, and minimal protocol overhead makes UDP particularly attractive for NTN, though it provides limited or no reliability guarantees.
On the other hand, \gls{tcp}, with a \gls{rto} of $200$ ms, has an application throughput similar to UDP for an altitude lower than $10\,000$ km, but its performance progressively deteriorates as the altitude increases, reaching only 5.6\% of the application rate at $30\,000$ km.
This effect is primarily due to the reliability mechanisms of \gls{tcp}, including ACKs, the three-way handshake connection procedure, packet retransmissions, and congestion control, which introduce significant delays.
Moreover, TCP timers, like the \gls{rto}, can expire given the long \gls{rtt} in NTN, causing the network to frequently go in slow start, with negative implications in terms of throughput.
Finally, sudden drops in the link quality, which may be common in satellite networks, can lead to frequent fluctuations in the congestion window, which prevents TCP from operating at full capacity.

To address these problems, the research community is exploring several solutions.
One approach is using Performance Enhancing Proxies (PEPs), which split the TCP connection into multiple segments to isolate and manage satellite links separately. Another one is to explore different TCP variants, including \gls{mptcp} to increase robustness through path diversity, as well as a cross-layer design where TCP adapts based on satellite dynamics and the conditions of the lower~layers.
Additionally, accurate fine-tuning of the TCP parameters, such as increasing the initial congestion window, adjusting RTO timers, and enabling selective ACKs,  is essential to accommodate the long propagation delays in~NTN.  

\section{Conclusions and future work}
\label{sec:conclusions}
\glspl{ntn} are emerging as a key enabler to extend 5G connectivity beyond terrestrial infrastructures. Specifically, while satellites offer unique coverage and connectivity advantages, their integration into the 5G ecosystem introduces significant technical challenges due to path loss, delay, and Doppler effects.
In this paper, we first provided an up-to-date overview of recent 3GPP 5G-NR NTN standardization activities to support satellite communication, from Rel. 17 to 20.
Then, we described key open challenges and possible protocol solutions for future NTN research, particularly regarding time and frequency synchronization and duplexing, resource allocation and retransmissions, coverage and mobility management, and network and transport layers.
Compared to the existing literature, where performance evaluations are often misaligned with standard protocol specifications, we used a system-level satellite network simulator, developed in ns-3, to test the effect of these challenges
on the full 5G NR-NTN protocol stack.
\balance
\bibliographystyle{IEEEtran}
\bibliography{biblio}

\onecolumn

\end{document}